\documentstyle[prl,aps,floats,epsf,color]{revtex}
\begin{document}
\baselineskip=12pt
\def\be{\begin{equation}}
\def\ee{\end{equation}}
\def\bea{\begin{eqnarray}}
\def\eea{\end{eqnarray}}
\def\E{{\rm e}}
\def\bearst{\begin{eqnarray*}}
\def\eearst{\end{eqnarray*}}
\def\peleven{\parbox{11cm}}
\def\peffec{\peight{\bearst\eearst}\hfill\peleven}
\def\pspace{\peight{\bearst\eearst}\hfill}
\def\ptwelve{\parbox{12cm}}
\def\peight{\parbox{8mm}}
\twocolumn
[\hsize\textwidth\columnwidth\hsize\csname@twocolumnfalse\endcsname

\title
{  Regeneration of Stochastic Processes: An Inverse Method }

\author
{ F. Ghasemi$^1$,  J. Peinke,$^2$, Muhammad Sahimi,$^3$ and  M.
Reza Rahimi Tabar,$^{1,4}$}

\vskip 1cm

\address
{ $^1$Department of Physics, Sharif University of Technology, P.O.
Box
11365-9161, Tehran 11365, Iran\\
$^2$Carl von Ossietzky University, Institute of Physics, D-26111
Oldenburg, Germany\\
$^4$Department of Chemical Engineering, University of Southern
California, Los Angeles, California 90089-1211, USA \\
$^3$CNRS UMR 6529, Observatoire de la C$\hat o$te d'Azur, BP 4229,
06304 Nice Cedex 4, France}
 \maketitle


\begin{abstract}

We propose a novel inverse method that utilizes a set of data to
construct a simple equation that governs the stochastic process
for which the data have been measured, hence enabling us to
reconstruct the stochastic process. As an example, we analyze the
stochasticity in the beat-to-beat fluctuations in the heart rates
of healthy subjects as well as those with congestive heart
failure. The inverse method provides a novel technique for
distinguishing the two classes of subjects in terms of a drift and
a diffusion coefficients which behave completely differently for
the two classes of subjects, hence potentially providing a novel
diagnostic tool for distinguishing healthy subjects from those
with congestive heart failure, even at the early stages of the
disease development.

\end{abstract}
\hspace{.3in}
\newpage
]

\noindent{\bf 1. Introduction}

Many natural or man-made phenomena, as well as the morphology of
numerous physical systems, are charactertized by a degree of
stochasticity. Turbulent flows, fluctuations in the stocks prices,
seismic recordings, the internet traffic, pressure fluctuations in
chemical reactors, and the surface roughness of many materials and
rock [1,2] are but a few examples of such phenomena and systems. A
long standing problem has been the development of an effective
reconstruction method for such phenomena. That is, given a set of
data for certain characteristics of a phenomenon, one would like
to develop an effective equation that can reproduce the data with
an accuracy comparable to the measured data. If such a method can
be developed, one may utilize it to, (1) reconstruct the original
process with similar statistical properties, and (2) understand
the nature and properties of the stochastic process.

In this paper we use a novel method to address this general
problem. The proposed method utilizes a set of data for a
phenomenon which contains a degree of stochasticity and constructs
a simple equation that governs the phenomenon. As we show below,
in addition to being highly accurate, the method is quite general;
it is capable of providing a rational explanation for complex
features of the phenomenon; it requires no scaling feature, and it
enables us to accomplish the tasks listed above. As an example, we
apply the method to analyze cardiac interbeat intervals which
normally fluctuate in a complex manner. We show that the
application of the method to the analysis of interbeat
fluctuations in the heart rates may potentially lead to a novel
method for distinguishing healthy subjects from those with
congestive heart failure (CHF).

\bigskip
\noindent{\bf 2. The Method}

\bigskip
We begin by describing the steps that lead to the development of a
stochastic equation, based on the (stochastic) data set, which is
then utilized to reconstruct the original data, as well as an
equation that describes the phenomenon.

(1) As the first step we check whether the data follow a Markov
chain and, if so, estimate the Markov time (length) scale $t_M$.
As is well-known, a given process with a degree of randomness or
stochasticity may have a finite or an infinite Markov time
(length) scale. The Markov time (length) scale is the minimum time
interval over which the data can be considered as a Markov process
[3-6]. To determine the Markov scale $t_M$, we note that a
complete characterization of the statistical properties of
stochastic fluctuations of a quantity $x$ in terms of a parameter
$t$ requires the evaluation of the joint probability distribution
function (PDF) $P_n(x_1,t_1;\cdots;x_n,t_n)$ for an arbitrary $n$,
the number of the data points. If the phenomenon is a Markov
process, an important simplification can be made, as the $n$-point
joint PDF, $P_n$, is generated by the product of the conditional
probabilities $p(x_{i+1},t_{i+1}|x_i,t_i)$, for $i=1,\cdots,n-1$.
A necessary condition for a stochastic phenomenon to be a Markov
process is that the Chapman-Kolmogorov (CK) equation [7],
\begin{equation}
p(x_2,t_2|x_1,t_1)=\int\hbox{d}(x_3)\;p(x_2,t_2|x_3,t_3)\;p
(x_3,t_3|x_1,t_1)\;,
\end{equation}
should hold for any value of $t_3$ in the interval $t_2<t_3<t_1$.
One should check the validity of the CK equation for different
$x_1$ by comparing the directly-evaluated conditional probability
distributions $p(x_2,t_2|x_1,t_1)$ with the ones calculated
according to right side of Eq. (1). The simplest way to determine
$t_M$ for stationary or homogeneous data is the numerical
calculation of the quantity,
$S=|p(x_2,t_2|x_1,t_1)-\int\hbox{d}x_3
p(x_2,t_2|x_3,t_3)\,p(x_3,t_3|x_1,t_1)|$, for given $x_1$ and
$x_2$, in terms of, for example, $t_3-t_1$ and considering the
possible errors in estimating $S$. Then, $t_M=t_3-t_1$ for that
value of $t_3-t_1$ for which $S$ vanishes or is nearly zero
(achieves a minimum).
\begin{figure}
\epsfxsize=7truecm\epsfbox{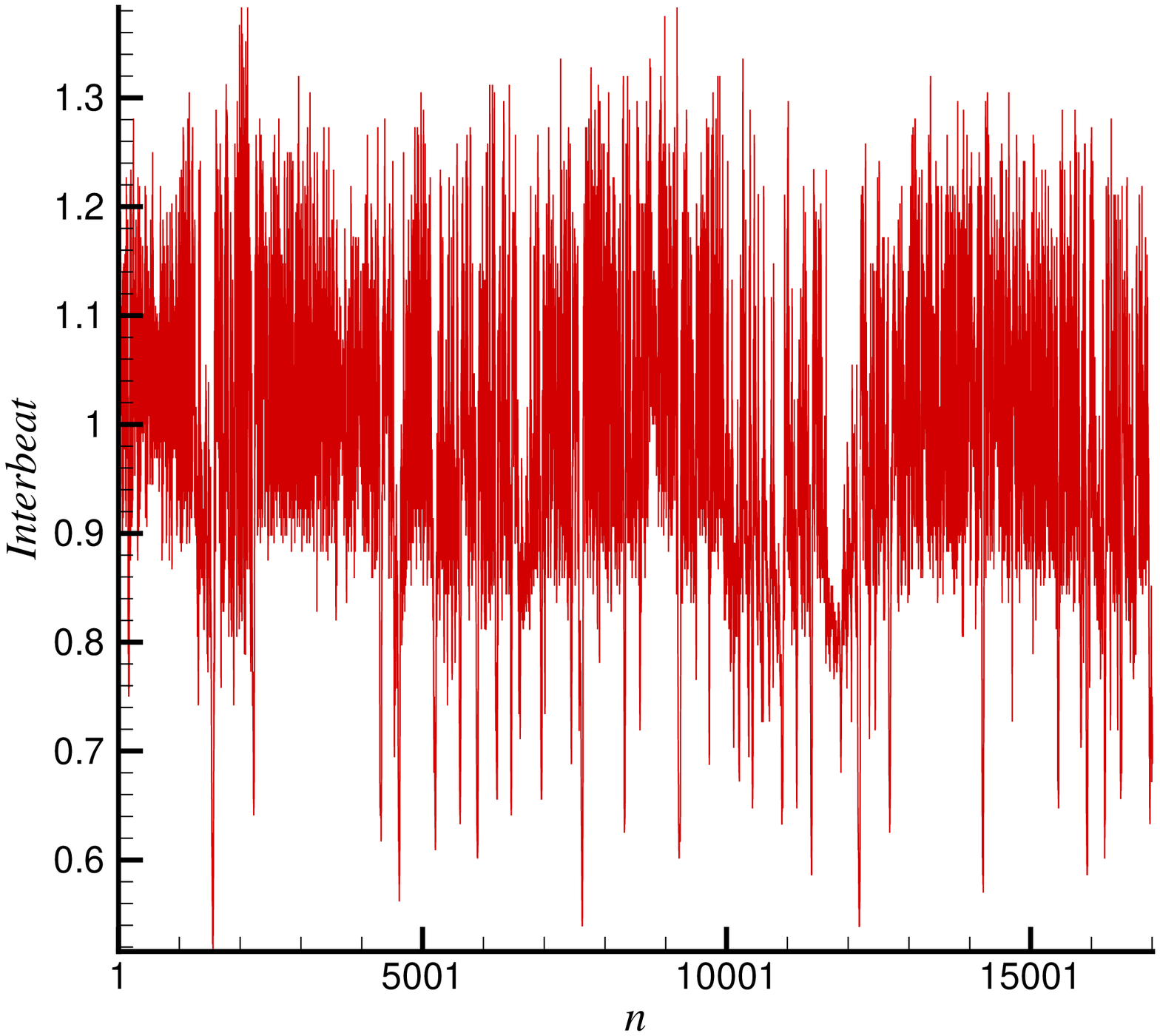}
\epsfxsize=7truecm\epsfbox{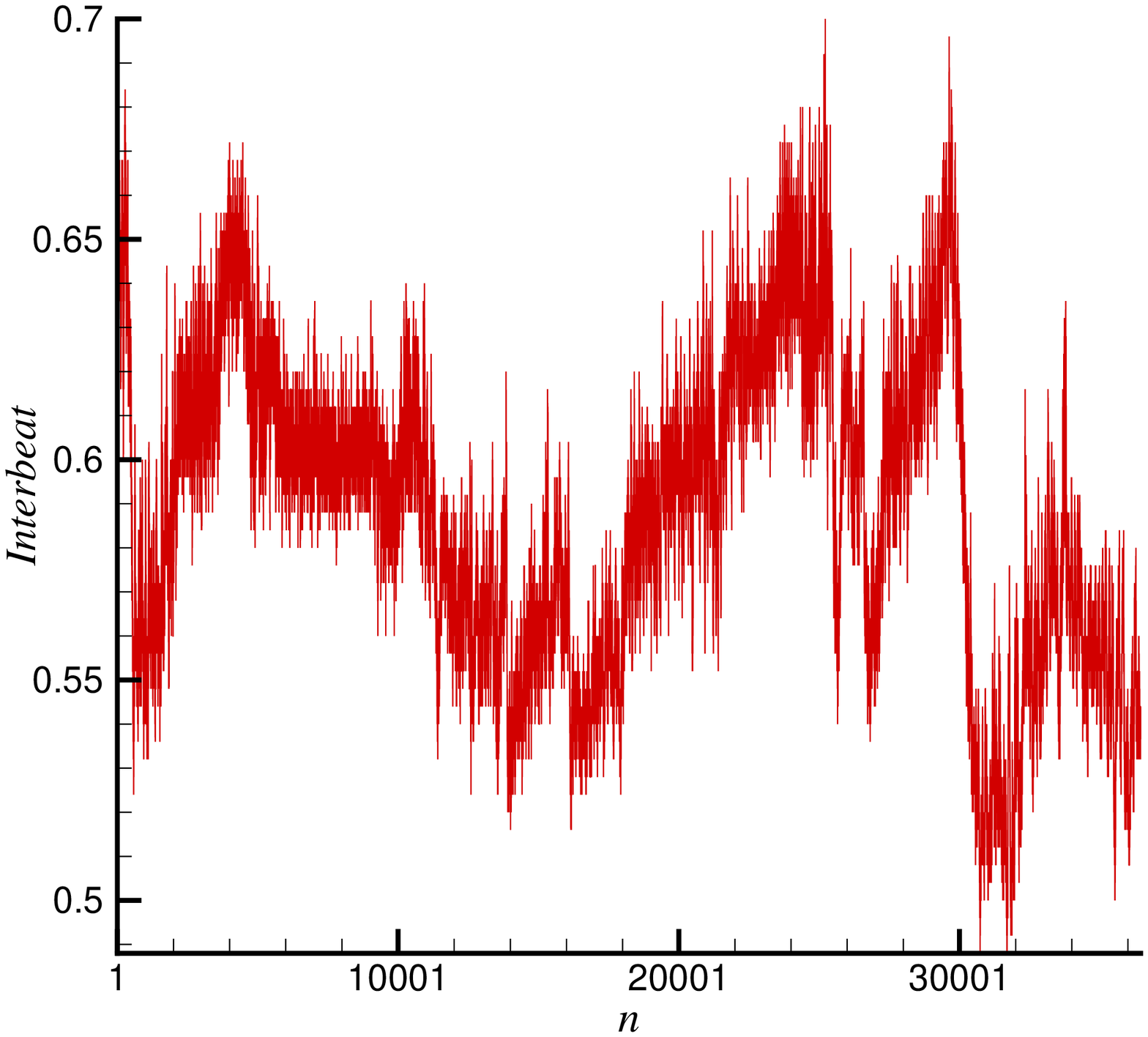}
 \narrowtext \caption{ Interbeats fluctuations of healthy subjects
(top), and those with congestive heart failure (bottom).}
 \end{figure}

(2) Deriving an effective stochastic equation that describes the
fluctuations of the quantity $x(t)$ constitutes the second step.
The CK equation yields an evolution equation for the change of the
distribution function $P(x,t)$ across the scales $t$. The CK
equation, when formulated in differential form, yields a master
equation which takes the form of a Fokker-Planck equation:
\begin{equation}
\frac{d}{dt}P(x,t)=\left[-\frac{\partial}{\partial x}D^{(1)}(x,t)+
\frac{\partial^2}{\partial x^2}D^{(2)}(x,t)\right]P(x,t)\;.
\end{equation}
The drift and diffusion coefficients, $D^{(1)}(x,t)$ and
$D^{(2)}(x,t)$, are estimated directly from the data and the
moments $M^{(k)}$ of the conditional probability distributions:
\begin{eqnarray}
&& D^{(k)}(x,t)=\frac{1}{k!}\hskip .2cm lim_{\Delta t\to
0}M^{(k)},
\cr\nonumber\\
&& M^{(k)}=\frac{1}{\Delta t}\int dx'(x'-x)^k p(x',t+\Delta
t|x,t).
\end{eqnarray}
We note that this Fokker-Planck equation is equivalent to the
following Langevin equation [8]:
\begin{equation}
\frac{d}{dt}x(t)=D^{(1)}(x)+ \sqrt{D^{(2)}(x)}\;\; f(t)\;,
\end{equation}
where $f(t)$ is a random force with zero mean and Gaussian
statistics, $\delta$-correlated in $t$, i.e., $\langle
f(t)f(t')\rangle=2\delta(t-t')$. Note that such a reconstruction
of a stochastic process does {\it not} imply that the data do not
contain any correlations, or that the above formulation ignores
the correlations.

(3) Regeneration of the stochastic process constitutes the last
step. Equation (4) enables us to regenerate a stochastic quantity
which is similar to the original one {\it in the statistical
sense}. The stochastic process is regenerated by iterating Eq. (4)
which yields a series of data {\it without memory}. To compare the
regenerated data with the original ones, we must take the spatial
(or temporal) interval in the numerical discretization of Eq. (4)
to be unity (or renormalize it to unity). However, the Markov
length or time is typically greater than unity. Therefore, we
should correlate the data over the Markov length or time scale.
There are a number of methods to correlate the generated data in
this interval [8-12]. Here, we propose a new technique which we
refer to as the {\it kernel} method, according to which one
considers a kernel function $K(u)$ that satisfies the condition
that,
\begin{equation}
\int_{-\infty}^\infty K(u)du=1\;,
\end{equation}
such that the data are determined by
\begin{equation}
x(t)=\frac{1}{nh}\sum_{i=1}^nx(t_i)K\left(\frac{t-t_i}{h}\right)\;,
\end{equation}
where $h$ is the window width. For example, one of the most useful
kernels is the standard normal density function,
$K(u)=(2\pi)^{-1/2}\exp(-\frac{1}{2} u^2)$. In essence, the kernel
method represents the data as a sum of `bumps' placed at the
observation points, with its function determining the shape of the
bumps, and its window width $h$ fixing their width. It is evident
that, over the scale $h$, the kernel method correlates the data to
each other.
\begin{figure}
\epsfxsize=7truecm\epsfbox{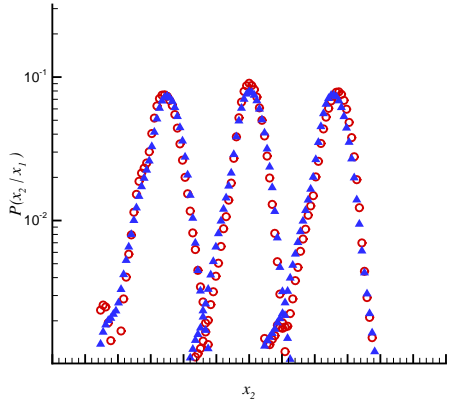} \narrowtext \caption{ Test of
Chapman-Kolmogorov equation for $x_1=-5$, $x_1=0$ and $x_1=5$. The
bold and open symbols represent, respectively, the
directly-evaluated PDF and the integrated PDF. The PDFs are
shifted in the vertical directions for better presentation. Values
of $x$ are measured in units of the standard deviation. }
 \end{figure}
\begin{figure}
\epsfxsize=7truecm\epsfbox{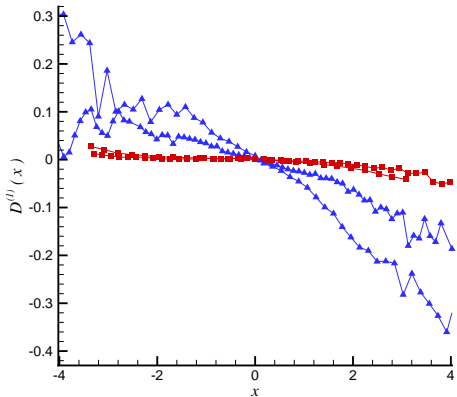}
\epsfxsize=7truecm\epsfbox{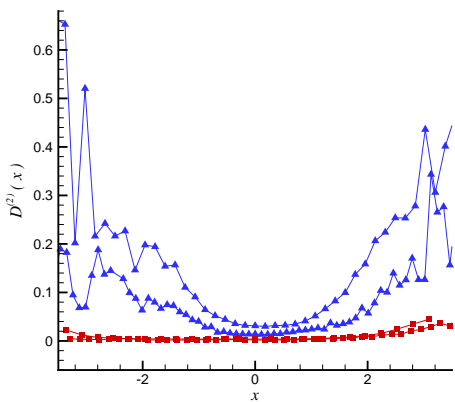} \narrowtext \caption{ The drift
and diffusion coefficients $D^{(1)}(x)$ and $D^{(2)}(x)$,
estimated by Eq. (3). For the healthy subjects (triangles)
$D^{(1)}(x)$ and $D^{(2)}(x)$ follow linear and quadratic behavior
in $x$, while for patients with CHF (squares) they follow third-
and fourth-order behavior in $x$.}
 \end{figure}
\begin{figure}
\epsfxsize=7truecm\epsfbox{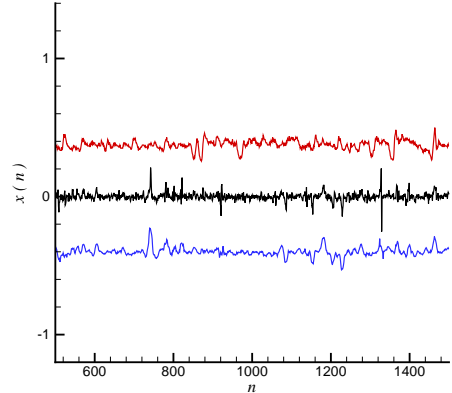} \narrowtext \caption{ The
curves show, from top to bottom, the actual interbeat data (for a
healthy subject), the regenerated data using the corresponding
Langevin equation, and the regenerated data using the kernel
method. The time series are shifted in the vertical directions for
better presentation.}
 \end{figure}
\vskip 3cm
\bigskip
\noindent{\bf 3. Application to Fluctuations in Human Heartbeats}

\bigskip
We now apply the above method to reconstruction of the
fluctuations in the human heartbeats of both healthy and ill
subjects by taking $h\simeq t_M$. Recent studies [13-18] reveal
that under normal conditions, beat-to-beat fluctuations in the
heart rate might display extended correlations of the type
typically exhibited by dynamical systems far from equilibrium. It
has been shown [14], for example, that the various stages of sleep
may be characterized by extended correlations of heart rates
separated by a large number of beats. We show that the Markov time
scale $t_M$, and the drift and diffusion coefficients of the
interbeat fluctuations of healthy subjects and patients with
congestive heart failure (CHF) have completely different
behaviour, when analyzed by the method we propose in this paper,
hence helping one to distinguish the two groups of the subjects.

\begin{figure}
\epsfxsize=7truecm\epsfbox{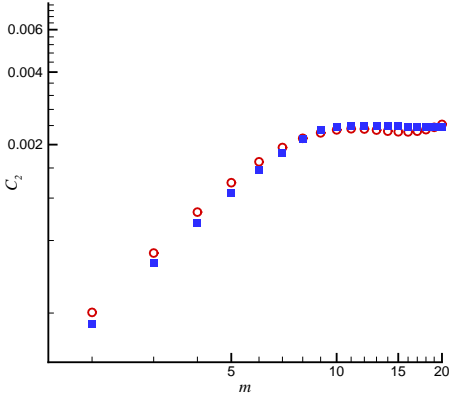} \narrowtext \caption{
Logarithmic plot of the second moment of the height-difference
versus $m$, for the actual data (circles) and the samples
regenerated by the kernel method (squares). The corresponding time
series are plotted in Fig. 4.}
 \end{figure}
We analyze both daytime (12:00 pm to 18:00 pm) and nighttime
(12:00 am to 6:00 am) heartbeat time series of healthy subjects,
and the daytime records of patients with CHF. Our data base
includes 10 healthy subjects (7 females and 3 males with ages
between 20 and 50, and an average age of 34.3 years), and 12
subjects with CHF, with 3 females and 9 males with ages between 22
and 71, and an average age of 60.8 years). Figure 1 presents the
typical data.

We first estimate the Markov time scale $t_M$ of the data for the
interbeat fluctuations. For the healthy subjects we find the
Markov time scale for the daytime data to be (all the values are
measured in units of the average time scale for the beat-to-beat
times of each subject), $t_M=3,3,3,1,2,3,3,2,3$ and 2. The
corresponding results for the nighttime records are, $t_M$ are
$3,3,1,3,3,2,3,3,2$ and $3$, respectively, comparable to those for
the daytime. On the other hand, for the daytime records of the
patients with CHF, the estimated Markov time scales are,
$t_M=151,258,760,542,231,257,864,8,366,393,385$, and 276.
Therefore, the healthy subjects have $t_M$ values that are much
smaller than those of the patients with CHF, hence providing an
unambiguous quantity for distinguishing the two.

We then check the validity of the CK equation for several $x_1$
triplets by comparing the directly-evaluated conditional
probability distributions $p(x_2,t_2|x_1,t_1)$ with the ones
calculated according to right side of Eq. (1). Here, $x$ is the
interbeat and for all the samples we define, $x\equiv (x-\bar
x)/\sigma$, where $\bar x$ and $\sigma$ are the mean and standard
deviations of the interbeats data. In Figure 2, the two
differently-computed PDFs are compared. Assuming the statistical
errors to be the square root of the number of events in each bin,
we find that the two PDFs are {\it statistically identical}.

The corresponding drift and diffusion coefficients $D^{(1)}(x)$
and $D^{(2)}(x)$ are displayed in Figure 3. We find that, in
addition to the Markov time scale $t_M$, the two coefficients
provide another important indicator for distinguishing the ill
from the healthy subjects: For the healthy subjects the drift
$D^{(1)}$ and the diffusion coefficient $D^{(2)}(x)$ are,
respectively, a linear and a quadratic function of $x$, whereas
the corresponding coefficients for patients with CHF follow a
third- and fourth-order equations in $x$. The analysis of the data
yields the following approximants for the healthy subjects,
\begin{eqnarray}
& & D^{(1)}(x)= -0.12 x\;, \cr \nonumber\\
& & D^{(2)}(x)=0.05 -0.042 x + 0.07 x^2\;,
\end{eqnarray}
whereas for the patients with CHF we find that,
\begin{eqnarray}
 D^{(1)}(x)&=&-0.0026 x - 0.0018 x^2 - 0.0007 x^3\;, \cr \nonumber \\
D^{(2)}(x)&=&0.0006 - 0.0007 x + 0.0005 x^2 \cr \nonumber \\
&+& 0.0003 x^3 + 0.0002x^4\;.
\end{eqnarray}

Equations (7) and (8) present the drift and diffusion coefficients
for a typical healthy subject and one with CHF. We note that the
final result for the Langevin equation is the same as the results
obtained in Ref. [18]. For other data measured for other patients
the functional dependence of $D^{(1)}$ and $D^{(2)}(x)$ would be
the same but with different numerical coefficients. The order of
magnitude of the coefficients is the same for all the healthy
subjects, and likewise for those with CHF (see also Ref. [19]).
Moreover, if we analyze different parts of the time series
separately, we find, (1) almost the same Markov time scale for
different parts of the time series, but with some differences in
the numerical values of the drift and diffusion coefficients, and
(2) that the drift and diffusion coefficients for different parts
of the time series have the same {\it functional forms}, but with
{\it different coefficients} in equations such as (7) and (8).
Hence, one can distinguish the data for sleeping times from those
for when the subjects are awake [20].

We also find another important difference between the heartbeat
dynamics of the two classes of subjects: Compared with the healthy
subjects, the drift and diffusion coefficients for the patients
with CHF are very small (reflecting, in some sense, the large
Markov time scale $t_M$). Hence, we suggest that one may use the
Markov time scales, the dependence of the drift and diffusion
coefficients on $x$, as well as their comparative magnitudes, for
characterizing the dynamics of human heartbeats and their
fluctuations, and to distinguish healthy subjects from those with
CHF. To our knowledge, this proposal is novel. Given its relative
simplicity, it would be most interesting to study whether this
proposal can be developed into a diagnostic tool for early
detection of congestive heart failure. Work in this direction is
in progress.

We compare in Figure 4 the original time series $x(n)$ with those
reconstructed by the Langevin equation [by, for example, using
Eqs. (4) and (7)] and the kernel method. While both methods
generate series that look similar to the original data, the kernel
method appears to better mimic the behavior of the original data.
To demonstrate the accuracy of Eq. (6), we compare in Figure 5 the
second moment of the stochastic function,
$C_2(m)=\langle[x(0)-x(m)]^2 \rangle$, for both the measured and
reconstructed data using the kernel method. The agreement between
the two is excellent. However, it is well-known that the agreement
between the second moments of a stochastic time series and its
reconstructed version is not sufficient for proving the accuracy
of the reconstruction method. Hence, we have also checked the
accuracy of the higher-order structure function,
$S_n=\langle|x(t_1)-x(t_2)|^n\rangle$ [21]. We find that the
agreement between $S_n$ for the original and reconstructed time
series for $n\leq 5$ is excellent, while the difference between
higher-order moments of the two times series, which are related to
the tails of the PDF of the $x-$increments, increases.

\bigskip
\noindent{\bf 4. Summary}

\bigskip
We have analyzed the interbeat fluctuations in the heart rates of
healthy subjects, as well as those with congestive heart failure,
by an inverse method for reconstruction of the stochastic process
that governs the fluctuations. The method, which is quite general
and can regenerate a stochastic process with high precision, is
based on utilizing measured data to estimate a drift and a
diffusion coefficients to be used in a Fokker-Planck, or an
equivalent Langevin, equation that describes the stochastic
process. The analysis of the times series for human heartbeat
dynamics using the new method, for both healthy subjects and those
with CHF, not only demonstrates the accuracy of the method, but
also potentially provides a novel technique for distinguishing the
heartbeat dynamics of the two classes of subjects.

We should point out that Stanley and co-workers [13,15-17,20,21]
analyze the type of data we considered in this paper by a method
different from what we present in the present paper. Their
analysis indicates that there may be long-range correlations in
the data, which might be characterized by self-affine fractal
distributions, such as the fractional Brownian motion or other
types of stochastic processes that give rise to such correlations.
They distinguish healthy subjects from those with CHF in terms of
the type of correlations that might exist in the data (negative as
opposed to positive correlations). The method proposed in the
present paper is different from that of Stanley and co-workers in
that, we analyze the data in terms of Markov processes. Although
our analysis does indicate the existence of correlations in the
data but, as is well-known in the theory of Markov processes, such
correlations, though extended, eventually decay. We distinguish
the healthy subjects from those with CHF in terms of the
differences between the drift and diffusion coefficients of the
Fokker-Plank equation which, in our view, provides a clearer and
more physical way of understanding the differences between the two
groups of the subjects. In addition, our method provides an
unambiguous way of reconstructing the data, hence providing a
means to predict the behavior of the data over periods of time
that are on the order of the Markov time scale. Although it
remains to be tested, we believe that our method is more sensitive
to small differences between the data for the two groups of the
subjects and, therefore, might eventually provide a diagnostic
tool for early detection of CHF in humans.

\bigskip
\noindent{\bf Acknowledgment}

\bigskip
We would like to thank Armin Bunde for useful comments on the
manuscript.

\end{document}